\documentclass[aps,prl,twocolumn,floatfix,superscriptaddress,10pt]{revtex4-1}
\usepackage{graphicx}
\usepackage{dcolumn}
\usepackage{bm}
\usepackage{color}
\usepackage{amsmath,amsfonts,amssymb,wasysym}
\usepackage{graphicx}
\usepackage[english]{babel}
\usepackage{color}
\usepackage{booktabs,longtable}
\usepackage{natbib}

\newcommand{\SDL}{\ell_{\rm sf}}
\newcommand{\SHA}{\theta_{\rm SHE}}
\newcommand{\SHAPt}{0.056\pm 0.010}
\newcommand{\gupdw}{g^{\uparrow \downarrow}}

\def\lsim{\mathrel{\rlap{\lower4pt\hbox{$\sim$}}\raise1pt\hbox{$<$}}}
\def\gsim{\mathrel{\rlap{\lower4pt\hbox{$\sim$}}\raise1pt\hbox{$>$}}}

\definecolor{myolive}{rgb}{0,0.5,0}

\newcommand{\beginsupplement}{%
        \setcounter{table}{0}
        \renewcommand{\thetable}{S\arabic{table}}%
        \setcounter{figure}{0}
        \renewcommand{\thefigure}{S\arabic{figure}}%
        \setcounter{equation}{0}
        \renewcommand*{\theequation}{S\arabic{equation}}
        \renewcommand{\citenumfont}{S}
        \renewcommand{\bibnumfmt}{S}
     }

\begin{document}

\author{J.-C.~Rojas-S\'anchez}
\affiliation{Unit\'e Mixte de Physique CNRS/Thales and Universit\'e Paris-Sud 11, 91767 Palaiseau, France}
\affiliation{INAC/SP2M, CEA-Universit\'e Joseph Fourier, F-38054 Grenoble, France}
\author{N.~Reyren}
\affiliation{Unit\'e Mixte de Physique CNRS/Thales and Universit\'e Paris-Sud 11, 91767 Palaiseau, France}
\author{P.~Laczkowski}
\affiliation{Unit\'e Mixte de Physique CNRS/Thales and Universit\'e Paris-Sud 11, 91767 Palaiseau, France}
\author{W.~Savero}
\affiliation{INAC/SP2M, CEA-Universit\'e Joseph Fourier, F-38054
Grenoble, France}
\author{J.-P.~Attan\'e}
\affiliation{INAC/SP2M, CEA-Universit\'e Joseph Fourier, F-38054 Grenoble, France}
\author{C.~Deranlot}
\affiliation{Unit\'e Mixte de Physique CNRS/Thales and Universit\'e Paris-Sud 11, 91767 Palaiseau, France}
\author{M.~Jamet}
\affiliation{INAC/SP2M, CEA-Universit\'e Joseph Fourier, F-38054 Grenoble, France}
\author{J.-M.~George}
\affiliation{Unit\'e Mixte de Physique CNRS/Thales and Universit\'e Paris-Sud 11, 91767 Palaiseau, France}
\author{L.~Vila}
\affiliation{INAC/SP2M, CEA-Universit\'e Joseph Fourier, F-38054 Grenoble, France}
\author{H.~Jaffr\`es}
\altaffiliation[Present address: ]{Peter Gr$\ddot{\rm u}$nberg Institut and Institute for Advanced Simulation, Forschungszentrum J$\ddot{\rm u}$lich and JARA, 52425 J$\ddot{\rm u}$lich, Germany}
\affiliation{Unit\'e Mixte de Physique CNRS/Thales and Universit\'e Paris-Sud 11, 91767 Palaiseau, France}

\title{Spin Pumping and Inverse Spin Hall Effect in Platinum: The Essential Role of Spin-Memory Loss at Metallic Interfaces.}

\date{\today}

\begin{abstract}
Through combined ferromagnetic resonance, spin-pumping and inverse spin Hall effect experiments in Co$|$Pt bilayers and Co$|$Cu$|$Pt trilayers, we demonstrate consistent values of $\SDL^{\rm Pt} = 3.4\pm 0.4$~nm and $\SHA^{\rm Pt} = \SHAPt$ for the respective spin diffusion length and spin Hall angle for Pt. Our data and model emphasizes the partial depolarization of the spin current at each interface due to spin-memory loss. Our model reconciles the previously published spin Hall angle values and explains the different scaling lengths for the ferromagnetic damping and the spin Hall effect induced voltage.
\end{abstract}

\maketitle

The direct control of the magnetization dynamics and magnetic damping \textit{via} spin-currents and spin-transfer torques is implemented in several magnetic nanoscale devices, as spin-torque magnetic random access memory and spin-torque nano-oscillators \cite{brataas2012,jungwirth2012}. Similar controls have recently been demonstrated using the spin-orbit related effect produced in ferromagnet-non magnetic metal layers with strong spin-orbit coupling (SOC)~\cite{miron2011,Liu2011a,Liu2012g,Liu2012,Hoffmann2013}. Spin-pumping~\cite{Mizukami2002,Tserkovnyak2002,Tserkovnyak2005,Jiao2013} is the method of choice to produce a spin-current in a SOC-material through ferromagnetic resonance (FMR) precession. It consists in generating an unbalanced chemical potential between the two spin channels (the so-called spin accumulation) from a metallic ferromagnet~\cite{Mizukami2002,Saitoh2006a,Ando2010a,Mosendz2010c,Ando2011,Azevedo2011,Feng2012,Ghosh2012,Nakayama2012,Shaw2012,Boone2013,Obstbaum2013,Vlaminck2013,Bai2013} or from a ferro/ferrimagnetic insulating oxide such as yttrium iron garnet (YIG)~\cite{Kajiwara2010,Sandweg2011,Kurebayashi2011,Jungfleisch2011,Wang2011,Chumak2012,Castel2012a,Althammer2013,OdAK2013,Hahn2013,Vlietstra2013}. The generated spin current is transformed into a charge current (or dc-voltage in an open-circuit) in the SOC-material by inverse spin Hall effect (ISHE). Large SOC can be found in 5\textit{d} or 4\textit{d} transition metal elements such as Pt~\cite{Kimura2007,Mosendz2010c,Ando2011,Azevedo2011,Liu2011a,Feng2012,Kondou2012,Nakayama2012,Castel2012a,Hahn2013,Vlaminck2013,Vlietstra2013,Bai2013}, $\beta$-Ta~\cite{Liu2012,Hahn2013}, $\beta$-W~\cite{Pai2012} or Pd~\cite{Mosendz2010c,Ando2010a,Ghosh2012,Shaw2012,Boone2013,Vlaminck2013}, as well as within heavy element alloys such as CuIr$_x$~\cite{Niimi2011} or CuBi$_x$~\cite{Niimi2012}, through \textit{intrinsic} or \textit{extrinsic} spin Hall effect with an overall efficiency given by the spin-Hall angle ($\SHA$). These combined techniques have also been employed to probe the spin-injection efficiency in group-IV semiconductors through a thin oxide barrier~\cite{Jain2012,Rojas-Sanchez2013,Pu2013}.

\begin{figure}
\includegraphics[width=8.5cm]{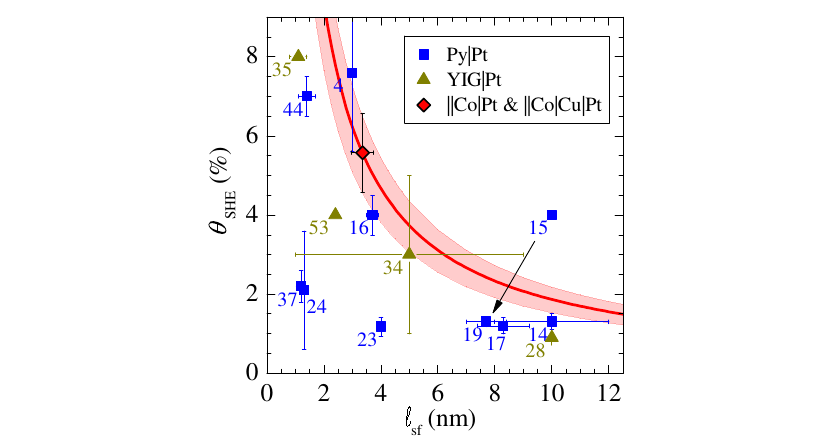}
\caption{Spin Hall angle $\SHA$ {\it vs.} spin diffusion length $\SDL$ for Pt films at room temperature. Py stands for Ni$_{80}$Fe$_{20}$. Data are extracted from literature, the number being the reference number.  The present work (Co$\vert$Pt and Co$\vert$Cu$\vert$Pt) includes the Pt thickness dependence by spin pumping and ISHE as well as in Refs.~\onlinecite{Azevedo2011,Feng2012,Nakayama2012}, or by STT-FMR in Ref.~\onlinecite{Liu2011,Kondou2012}. In the other studies, either the value of $\SDL$ or the value of $\SHA$ was adopted. The arrow shows the correction of the parameters from Ref.~\onlinecite{Ando2011} to Ref.~\onlinecite{Nakayama2012} as the Pt thickness-dependence is performed. The thick line represents a constant product $\SDL^{\rm Pt}{\rm (nm)}\cdot \SHA^{\rm Pt}{\rm (\%)}=18.8$.}
\label{Fig1}
\end{figure}

The case of bulk Pt is particularly interesting from a fundamental point of view, as well as for technological applications~\cite{jungwirth2012}. However, published values of both the spin-diffusion length ($\SDL$) and $\SHA$ for Pt are scattered over one order of magnitude, ranging from 1 to 10~nm for $\SDL$ and from 0.01 to 0.08 for $\SHA$ (Fig.~\ref{Fig1}). Note that these values are measured in thin multilayers, {\it i.e.} systems in which interfaces play a dominant role. The dispersion is also explained by the correlation between $\SDL$ and $\SHA$ in the expression of the charge-current generated by spin-pumping ({\it e.g.} in Eq.\ref{eqIC}).

In this letter, we attempt to reconcile the published Pt data. We emphasize the central role of the unavoidable spin relaxation known as spin memory loss (SML) at $3d|5d$ interfaces, here Co$\vert$Pt and Co$\vert$Cu$\vert$Pt where a spin-current is generated by FMR methods. We develop a model to extract reliable values by taking into account the SML. By using complementary data of FMR and ISHE in the microwave regime for different thicknesses of Pt, we succeeded to disentangle both $\SHA$ and $\SDL$. On one hand, FMR analysis gives access to the effective damping parameter $\alpha$ which is sensitive to the total dissipated transverse spin-current. On the other hand, the ISHE signal probes only the spin-current absorbed in the bulk part of the SOC-material ({\it i.e.} Pt in our case). Consequently, the thickness dependences of $\alpha$ and the ISHE signal in SOC-material scale respectively with the interfacial layer and the $\SDL$. The main goal of this letter is to demonstrate that neglecting the spin-current absorbed at the interfaces leads to an incorrect estimation of the $\SHA$ of Pt. Our method allows solid values of $\SDL^{\rm Pt}=3.4\pm 0.4$~nm and $\SHA^{\rm Pt}=\SHAPt$ for Pt to be determined and may reconcile the general trend of published data.

\begin{figure}
\includegraphics[width=8.5cm]{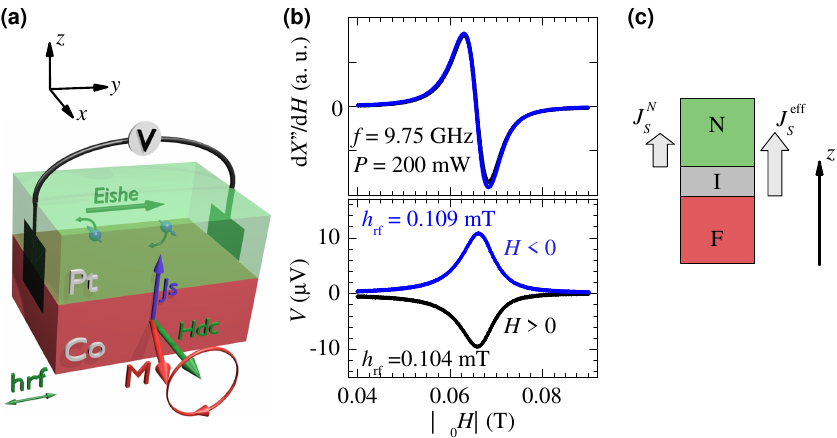}
\caption{(a) Schematic of the principle of spin-pumping-ISHE experiments in the case of $\|$Co$\vert$Pt bilayers. (b) Typical measurements of FMR spectrum (top) and ISHE voltage (bottom) for a $\|$Co(15)$\vert$Pt(10) sample. (c) Model of a trilayer system with spin memory loss (SML): F$\vert$I$\vert$N represent the ferromagnetic layer, the interface region and the SOC-material. $J_{\rm S}^{\rm eff}$ and $J_{\rm S}^{\rm N}$ are the effective spin-current emitted from the ferromagnet and the spin-current that reaches the SOC-material. Taking into account the SML at the interface implies $J_{\rm S}^{\rm N}<J_{\rm S}^{\rm eff}$.}
\label{Fig2}
\end{figure}

We deposited a series of $\|$Co(15)$\vert$Pt($t$) and $\|$Co(15)$\vert$Cu(5)$\vert$Pt($t$), varying the thickness $t$ of Pt, the numbers in the bracket indicate the thickness in nanometers and $\|$ the position of the substrate. The samples were grown by magnetron sputtering in a single deposition chamber on SiO$_2$-terminated Si wafers. Samples are then cut in an elongated rectangular shape of typical dimension $L\times W = 2.4\times 0.4$~mm$^2$. Combined FMR and ISHE measurements were performed at room temperature in a split-cylinder microwave resonant cavity. The rf magnetic field $h_{\rm rf}$ is along the long axis and the external applied dc magnetic field $H_{\rm dc}$ along the width of the rectangle [Fig.~\ref{Fig2}(a)]. The frequency of $h_{\rm rf}$ is fixed at $9.75$~GHz whereas $H_{\rm dc}$ is swept through the FMR condition. The amplitude of $h_{\rm rf}$ was determined by measuring the $Q$ factor of the resonant cavity with the sample placed inside, for each measurement. The derivative of FMR energy loss is measured at the same time as the voltage taken across the long extremity of the sample. We have also carried out a frequency dependence ($3-24$~GHz) of the FMR spectrum in order to determine the effective saturation magnetization $M_{\rm eff}$ as well as the damping constant $\alpha$. Details of such calculations are found in the supplemental material (SM). For damping analysis, we needed a reference sample free of spin-current dissipation, {\it i.e.} without SML. Ideally one would use a single Co layer, but to prevent its oxidation, we grew a capping layer of Al.

\begin{figure}
\includegraphics[width=8.5cm]{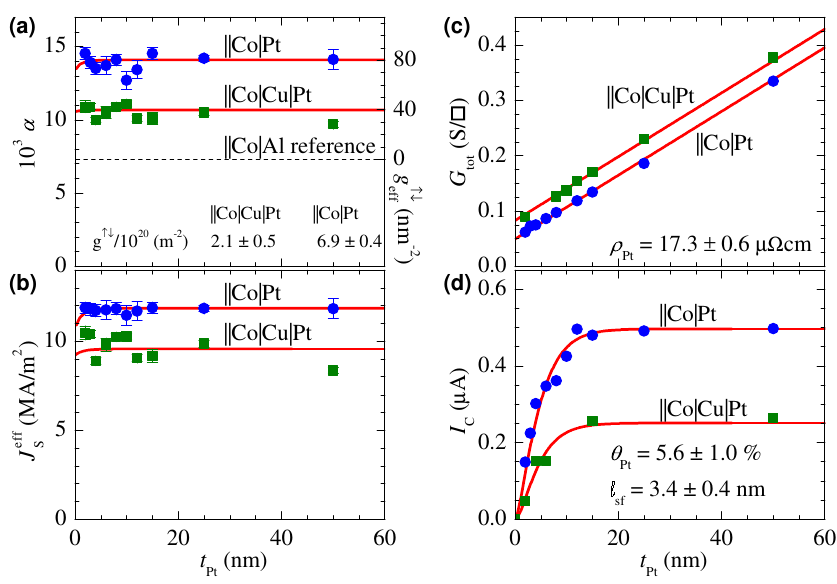}
\caption{(Color online) Platinum thickness dependence of the spin-injection parameters for $\|$Co(15)$\vert$Cu(5)$\vert$Pt$(t)$ (\color{myolive}$\blacksquare$\color{black}) and $\|$Co(15)$\vert$Pt$(t)$ (\color{blue}$\CIRCLE$\color{black}) systems at room temperature. (a) Damping constants (after frequency dependence) and their fits with Eq.~(\ref{eqDalfa}), (b) Effective spin currents, Eq.~(\ref{eqJseff}), and their fit proportional to $\gupdw_{\rm eff}$ in Eq.~(\ref{Eqgeff}) (1\,MA/m$^2$ corresponds to 0.33\,nJ/m$^2$), (c) Sheet conductances $G_{\rm tot}$ of the full stack multilayers and their linear fits, and (d) Charge currents and the fits with Eq.~(\ref{eqIC}). In order to fit $I_{\rm C}$ for the $\|$Co$\vert$Cu$\vert$Pt ($\|$Co$\vert$Pt) system, we used $r_{\rm sI}=0.85~(1.7)$~f$\Omega\,$m$^2$, and SML $\delta=1.2~(0.9)$. }
\label{Fig3}
\end{figure}

Raw data of a typical FMR spectrum and ISHE voltage measurements performed simultaneously on a $\|$Co(15)$\vert$Pt(10) sample are shown in Fig.~\ref{Fig2}(b). $H_{\rm dc}$ is parallel to the film plane along the $x$-axis (black data), and when the sample is turned 180$^\circ$ around the $y$ axis (blue data) the ISHE voltage is reversed. As expected, we observe that both ISHE voltage curves have their peak at the resonance field of the FMR spectrum with the same linewidth \cite{Mosendz2010c,Ando2011,Azevedo2011}. In order to calculate the charge current $I_{\rm C}$ we measured directly the sheet conductance ($G_{\rm tot}$) of the full stack by a four probe method. It follows that $I_{\rm C}=V_{\rm ISHE}\frac{W}{L}G_{\rm tot}$, where $V_{\rm ISHE}$ is the average weighted by factor $h_{\rm rf}^2$ of the Lorentzian amplitudes of the fitted voltage data. The sheet conductance $G_{{\rm Co}\vert{\rm Pt}}$ of the $\|$Co(15)$\vert$Pt($t$) bilayers as a function the Pt thickness $t$ is displayed in Fig.~\ref{Fig3}(c). The perfectly linear behavior indicates a {\em thickness independent} Pt bulk resistivity of $17.9\pm 0.2~\mu\Omega\,$cm (at room temperature) down to $2$~nm. The same conclusions can be raised for the $\|$Co(15)$\vert$Cu(5)$\vert$Pt($t$) trilayer series [Fig.\ref{Fig3}(c)], giving a similar Pt resistivity of $16.7\pm 0.2~\mu\Omega\,$cm. Importantly this proves that inserting the Cu(5) layer does not impact significantly the Pt layer quality. The same method of sheet conductance analysis is applied to Co and gives a bulk resistivity for Co of about $\rho_{\rm Co}=17~\mu\Omega\,$cm (see SM) leading to a characteristic Co spin-resistance of about $r_{\rm sF}=\rho_{\rm Co}\times \SDL^{\rm Co}\approx 6.7~{\rm f}\Omega\,{\rm m}^2$ at room temperature with a typical $\SDL^{\rm Co}$ of $38\pm 12$~nm \cite{Piraux1998}.

We will now focus on FMR and ISHE experimental data obtained on $\|$Co(15)$\vert$Cu(5)$\vert$Pt($t$) trilayer series, free of induced magnetic moments in Pt. In that sense, this series plays the role of a reference in which pure interfacial SML effects can be analyzed. The magnetic damping parameter $\alpha$ as a function of Pt thickness $t$, ranging from $2$~nm to $50$~nm is displayed in Fig.~\ref{Fig3}(a). We measure an almost thickness-independent $\alpha$ parameter down to $2$~nm of Pt, having a measured value close to $\alpha_{\rm Co|Cu|Pt}=10.5\times 10^{-3}$ in the whole Pt thickness range whereas the corresponding $\alpha_{\rm Co}$ for Co was measured at $7.56\times 10^{-3}$ in the  $\|$Co(15)$\vert$Al(7) reference sample, free of spin current-dissipation.

The combination of spin-pumping and ISHE results in the expression for the effective spin-current density ($J_{\rm S}^{\rm eff}$) pumped outward the ferromagnet and collected in the SOC-metal according to~\cite{Ando2011,Mosendz2010c,Azevedo2011}:
\begin{equation}\label{eqJseff}
J_S^{\rm eff}  =  \frac{2e}{\hbar} \cdot \frac{\gupdw_{\rm
eff}\gamma^2\hbar h_{\rm rf}^2}{8\pi \alpha_{\rm }^2} \cdot
\frac{4\pi M_{\rm eff}\gamma  + \sqrt{\left( 4\pi M_{\rm
eff}\gamma\right)^2 + 4\omega ^2}} {\left( 4\pi M_{\rm eff}\gamma
\right)^2 + 4\omega ^2} ,
\end{equation}
where $\omega=2\pi f$ is the microwave pulsation, $e$ is the electron charge, $\hbar$ is the reduced Planck constant, $\gamma=g \mu_{\rm B} / \hbar$ is the gyromagnetic ratio with $g$ the Land\'e factor and $\mu_B$ the Bohr magnetron. 
The enhancement of the magnetic damping in multilayers is assigned to the spin-current dissipation by spin-pumping mechanisms \cite{Mizukami2002,Tserkovnyak2002}. The damping parameter varies over a very short lengthscale, smaller than $2$~nm, due to the total spin-current dissipation. The enhancement of $\alpha$ is generally related to the effective spin mixing conductance $\gupdw_{\rm eff}$ by the following relation~\cite{Tserkovnyak2005,Azevedo2011,Jiao2013}: 
\begin{equation}
\Delta\alpha = \frac{g \mu_{\rm B}}{4\pi M_{\rm eff} t_{\rm Co}} \,\gupdw_{\rm eff} ,\\
\label{eqDalfa}
\end{equation}
where $t_{\rm Co}$ is the Co thickness. Note that such $\gupdw_{\rm eff}$ effective quantity describes the total spin-current dissipated outward Co itself and that it contains self-consistently the spin-backflow contribution. The effective spin-mixing conductance at saturation is then estimated at $\gupdw_{\rm eff, Co|Cu|Pt}\approx 40$~nm$^{-2}$.

What about the ISHE data measured on the same series of samples? The ISHE current $I_{\rm C}$ flowing in ``bulk'' Pt \textit{vs.} Pt thickness $t$ is displayed in Fig.~\ref{Fig3}(d). The corresponding variation for $I_{\rm C}$ can be described by the conventional $I_{\rm C}=W\SDL^{\rm N}\SHA\tanh\left(\frac{t_{\rm N}}{2\SDL^{\rm N}}\right)J_{\rm S}^{\rm eff}$ function ($t_{\rm N}$ is the thickness of the SOC-material), with a characteristic lengthscale $\SDL^{\rm N}$ of about $3.4$~nm, identified here as the intrinsic $\SDL^{\rm Pt}$. At this point we conclude that the magnetic damping and the ISHE current occur over two different lengthscales, related to either the interface or the bulk properties of the SOC-material: the total spin-current is dissipated through the enhancement of $\Delta \alpha$ over less than $2$~nm, while the spin current is absorbed over about $3.4$~nm in Pt.

The standard bilayer approach fails to describe our measurements in two aspects. First the increase of the magnetic damping $\Delta \alpha$ should scale as $I_{\rm C}$, which is not the case. Non-linearities between spin-current and damping enhancement reported in interfaces with an insulating oxide \cite{Wang2011} cannot be invoked because they are not observed in metallic multilayers for spin current densities in the range used in FMR \cite{Ando2009}. Secondly, if we compare the data for the $\|$Co$|$Cu$|$Pt and the $\|$Co$|$Pt series (shown in Fig.~\ref{Fig3}), the insertion of a thin $5$~nm Cu layer in between Co and Pt should have no impact on the extracted value of $\SHA^{\rm Pt}$, because the Cu thickness is much smaller than its own $\SDL$. However, with the conventional model, we have estimated $\SHA$ that changes by a factor of 2 when the Cu layer is inserted. Therefore the conventional extraction method from the $I_{\rm C}$ \textit{vs.} Pt thickness variation cannot explain this difference. Note that our previous sheet conductance measurements demonstrate that a change of the material properties of Pt  with thickness, cannot be invoked to explain such discrepancy.

Examining now the case of $\|$Co(15)$\vert$Pt($t$) bilayer series (Fig.~\ref{Fig3}), we draw the same qualitative conclusions than previously: There are two different lengthscales for the Pt thickness dependence of the magnetic damping $\Delta \alpha$ ($<2$~nm) and of the ISHE current $I_{\rm C}$ ($\sim 4$~nm). The damping $\alpha$ was measured at a level of $\alpha=14\times 10^{-3}$. This manifests an effective spin-mixing conductance $\gupdw_{\rm eff, Co|Pt}\approx 80$~nm$^{-2}$ twice as large as in the trilayer. However the evidence of two different lengthscales in the $\|$Co(15)$\vert$Pt($t$) series may find its origin in an other phenomenon, namely the induced polarization in Pt. In this scenario, the transverse spin-current would dissipate by spin decoherence due to magnetic moments induced in the first atomic layers of Pt in contact with Co (proximity effects)~\cite{Ghosh2013}. Nevertheless, even in that case, questions persist concerning the exact mechanism for interfacial spin decoherence. Taking into account the strong coupling of these magnetic moments in Pt with the Co magnetization, an overall effective spin-mixing conductance would be insensitive to interface spin dissipations as reported by Tserkovniak \textit{et al}~\cite{Tserkovnyak2005}.

From spin-transport and magnetoresistance experiments on metallic multilayers, it is well established that metallic interfaces dissipate spin-current by SML~\cite{Eid2002} mainly due to interfacial diffusion and disorder, in particular for transition metal 3$d|$Pt interfaces such as Cu$\vert$Pt~\cite{Kurt2002} and Co$\vert$Pt~\cite{Nguyen2013}. The physical parameter governing such SML processes is given by the spin-flip parameter $\delta=t_\textrm{I}/\SDL^{\textrm{I}}$ which can be viewed as the ratio between the effective interface ``thickness'' $t_{\rm I}$ and the interface spin diffusion length $\SDL^{\rm I}$, which becomes short with disorder. SML generally results in a large $\delta$ measured at low temperatures: $\delta=0.25$ for Co$|$Cu \cite{Eid2002} and $0.9$ for both Cu$|$Pt \cite{Kurt2002} and Co$|$Pt \cite{Nguyen2013}, corresponding respectively to a probability of the depolarization ($1-\exp(-\delta)$) of 22\% and 60\%. By comparison with Pt, SML at Cu$|$Pd interface is only limited by $\delta=0.25$ (20\% of SML probability) \cite{Sharma2007}, which means that the standard bilayer experimental analysis for Pd should be more reliable \cite{Mosendz2010c,Ando2010a,Ghosh2012,Shaw2012,Boone2013,Vlaminck2013}. Therefore 3$d|$Pt interfaces require a trilayer analysis taking into account an interfacial layer to describe transport, relaxation, and diffusion of the spin-current generated by spin-pumping as displayed in Fig.~\ref{Fig2}(c). In that picture, the interfacial spin-resistance $r_{\rm sI}$ equals $r_{\rm b}/\delta$, where the $r_{\rm b}$ is the interface resistance. If the SML is large, the spin-current will be mainly dissipated in the interfacial layer. As a consequence, $\Delta \alpha$ increases \textit{without} creating a charge current $I_{\rm C}$ in the bulk SOC-material by ISHE. This is what we observe.

Taking into account such an interfacial layer in the expression for the spin-current injected from Co ($J_{\rm S}^{\rm eff}$) and absorbed in the bulk SOC-material ($J_{\rm S}^{\rm N}$), one gets (see SM) for the ratio $R_{\rm SML}$ between the spin-current absorbed in bulk SOC-material (Pt) and the total spin-current dissipated (interface+bulk):
\begin{equation}
R_{\rm SML} = \frac{J_{\rm S}^{\rm N}}{J_{\rm S}^{\rm eff}} = \frac{r_{\rm sI}}{r_{\rm sI}\cosh(\delta)+r_{\rm sN}\sinh(\delta)} ,
\label{EqRSML}
\end{equation}
where $r_{\rm sN}=r_{\rm sN}^{\infty}\coth\left(t_{\rm sN}/\SDL^{\rm N}\right)$ stands for the spin-resistance of the SOC-material of finite thickness $t_{\rm N}$. Note that the ratio $R_{\rm SML}$ does not depend on the rate of backflow and this will make our conclusion very robust. In our systems, the Pt spin-resistance is $r_{\rm sN}^{\infty}=0.58$~f$\Omega\,$m$^2$. The interface resistances at room temperature are unknown in our systems but typical values reported at 4.2~K are $2AR=1.5$\,f$\Omega$\,m$^2$ ($r_{\rm sI}=1.7$\,f$\Omega$\,m$^2$) for Cu$|$Pt \cite{Kurt2002} and $2AR^\star=1.0$\,f$\Omega$\,m$^2$ ($r_{\rm sI}=2.0$\,f$\Omega$\,m$^2$) for Co$|$Cu \cite{Sharma2007,Bass2007}, resulting in an effective Co$|$Cu$|$Pt spin resistance of 0.85~f$\Omega$\,m$^2$ (see SM). We expect these values to be the lower bounds for the room temperature values. For large values of interfacial $\delta$, the variation of $J_{\rm S}^{\rm eff}$ (and then $\Delta \alpha$) is on the scale of $t_{\rm I}$ whereas the one of $J_{\rm S}^{\rm N}$ is on the scale of the inverse of $r_{\rm sN}$ that is $\SDL^{\rm N}$, as observed in our experiments. Finally the corrected expression for charge current is
\begin{equation}\label{eqIC}
{{I}_{\text{C}}}=\frac{\theta
_{\text{SHE}}^{\text{N}}\ell_{\text{sf}}^{\rm N}W J_{\text{S}}^{\text{eff}}\tanh {{\left( \frac{{{t}_{\text{N}}}}{2\ell
_{\text{sf}}^{\text{N}}}
\right)}_{{}}}{{r}_{\text{sI}}}}{{{r}_{\text{sI}}}\cosh \left( \delta \right)+r_{\text{sN}}^{\infty }\coth \left(
\frac{{{t}_{\text{N}}}}{\ell _{\text{sf}}^{\text{N}}} \right)\sinh \left( \delta \right)} ,
\end{equation}
and the corrected effective spin mixing conductance $\gupdw_{\rm eff}$ is written as (see SM):
\begin{widetext}
\begin{equation}
g_{\text{eff}}^{\uparrow \downarrow }={{g}^{\uparrow \downarrow }}\frac{{{r}_{\text{sI}}}\cosh \left(\delta \right)+r_{\text{sN}}^{\infty }\coth \left( \frac{{{t}_{\text{N}}}}{\ell _{\text{sf}}^{\text{N}}} \right)\sinh \left( \delta \right)}{{{r}_{\text{sI}}}\left[ 1+\frac{1}{2}\sqrt{\frac{3}{\varepsilon }}\coth \left( \frac{{{t}_{\text{N}}}}{\ell _{\text{sf}}^{\text{N}}} \right) \right]\cosh \left( \delta \right)+\left[ r_{\text{sN}}^{\infty }\coth \left( \frac{{{t}_{\text{N}}}}{\ell _{\text{sf}}^{\text{N}}} \right)+\frac{1}{2}\frac{{{r}_{\text{sI}}}^{2}}{r_{\text{sN}}^{\infty }}\sqrt{\frac{3}{\varepsilon }} \right]\sinh \left(\delta \right)} ,
\label{Eqgeff}
\end{equation}
\end{widetext}
where $\varepsilon=\tau_{\rm el}/\tau_{\rm sf}^{\rm N}=0.1$ is the ratio of the spin-conserved to spin-flip relaxation times, with $\varepsilon=0.1$ for Platinum~\cite{Tserkovnyak2005,Azevedo2011,Nakayama2012,Jiao2013}. 

We discuss now the different issues of the quantitative analyses of the FMR-ISHE extracted using the bilayer treatment method conventionally used in the literature:

({\it i}\,) The bilayer analysis generally gives a shorter $\SDL^{\rm N}$ than the real one considering only the variation lengthscale of the parameter $\alpha$ (Note that it also applies to the case of the STT-FMR technique \cite{Liu2011a,Liu2011,Kondou2012}). The damping is more related to the interfacial SML. 

({\it ii}\,) The level of the spin-current penetrating into the bulk SOC-material is smaller by a ratio $R_{\rm SML}$ than the one given from $J_{\rm S}^{\rm eff}$, leading to a $\SHA$ under-estimated by the same ratio, $R_{\rm SML}$, if the interfaces are assumed to be transparent. 

The trilayer analysis allows us to fit consistently all the experimental FMR and ISHE data (Fig.~\ref{Fig3}), and gives a value of $\SDL^{\rm Pt}=3.4\pm0.4$~nm for bulk Pt ($r_{\rm sPt}^{\infty}=0.58$~f$\Omega\,$m$^2$), and $\SHA^{\rm Pt}=\SHAPt$, with the following values $r_{\rm sI}({\rm Co|Cu|Pt})=0.85$~f$\Omega$\,m$^2$, $r_{\rm sI}({\rm Co|Pt})=0.83$~f$\Omega$\,m$^2$ and the aforementioned corresponding $\delta$ parameters (Fig.~\ref{Fig3}). This corresponds to an intrinsic spin Hall conductivity $\sigma_{\rm SHE}$ of $3.2\times 10^{3} (\Omega\,{\rm cm})^{-1}$ for Pt. Note that the estimation of the error on $\SHA^{\rm Pt}$ is not only the statistical error related to the fit (0.1\%), but is mainly due to the uncertainties for the values for $r_b$ and $\delta$ reported in the literature. Interestingly, one can notice that the specific value of $3.4$~nm for $\SDL^{\rm Pt}$ is in agreement with the value given by different works \cite{Nguyen2013} for this level of resistivity $\rho_{\rm Pt}=17.3~\mu\Omega\,$cm. The thick red line in Fig.~\ref{Fig1} corresponds to a constant $\SDL^{\rm Pt}\cdot\SHA^{\rm Pt}$ product [see Eq.~(\ref{eqIC})]. Finally we emphasize that SML is not restrained to the spin-pumping experiments but applies to all spin current injection phenomena. For example, strong SML at Pt$|$Cu interface (compared to Pt$|$YIG) could also explain the strong reduction of spin Hall magnetoresistance signal at Pt$|$Cu$|$YIG with respect to Pt$|$YIG \cite{Nakayama2013}.

In conclusion the present work demonstrates that the spin memory loss at 3$d$ transition metal$|$Pt interfaces induces a strong interfacial depolarization of the spin-current injected in Pt by spin-pumping methods. Such spin-current depolarization largely affects the ability to correctly extract both spin diffusion length and spin Hall angle in Pt, and hence requires a careful treatment by considering a more complete trilayer spin-current diffusion/relaxation model. This interfacial SML effects need to be carefully addressed for the design of efficient devices using SHE, for example in order to control magnetization reversal. In particular, spin memory loss in future devices can be reduced by interface engineering using multilayers with smaller SML.

\begin{acknowledgments}
We acknowledge U. Ebels, W. E. Bailey, S. Gambarelli, and G. Desfonds for technical support with the FMR measurements, J. Bass, A. Fert and F. Freimuth for fruitful discussions, and A. S. Jenkins for a careful reading the manuscript. This Work was partly supported by the French Agence Nationale de la Recherche (ANR) through projects SPINHALL (2010-2013) and SOSPIN (2013-2016).
\end{acknowledgments}

%


\beginsupplement
\clearpage
\onecolumngrid
\section{Supplemental Material\\  Spin Pumping and Inverse Spin Hall Effect in Platinum:\\ The Essential Role of Spin-Memory Loss at Metallic Interfaces}

\paragraph{}
We demonstrate the equations given in the main text. We show details of damping calculation after frequency dependence of FMR spectrum and the experimental and details about Co and Pt resistivities.

\date{\today}

\maketitle

\beginsupplement
\onecolumngrid


\subsection*{Calculation of the spin current density profile at different layers and interfaces }

Following the standard Valet-Fert diffusion model \cite{SupValet1993}, the steady-state transverse spin electro-chemical potential can be
expressed in the following form (with $\mu_s=\mu_\uparrow-\mu_\downarrow$) \cite{SupTakahashi2003}:
\begin{equation}
\nabla^{2}\mu_{\rm s}=\frac{\mu_{\rm s}}{\SDL^{2}}   \quad,   \label{eq1}
\end{equation}
where $\ell_{sf}$ is the spin diffusion length. The dc spin current writes as:
\begin{eqnarray}
j_{\rm s}^{\rm eff} & = & -\frac{\hbar}{2e^{2}}\frac{1}{\rho}\nabla\mu_{\rm s}\nonumber\\
j_{\rm s} & = & -\frac{1}{e\rho}\nabla\mu_{\rm
s}=-\frac{1}{e}\frac{\SDL}{r_{\rm s}}\nabla\mu_{\rm s}   \quad,   \label{eq:js}
\end{eqnarray}
where $\rho$ is the resistivity and $\hbar/2e^{2}=2054\,\Omega$. Here $j_{\rm s}^{\rm eff}$ is expressed in units of J/m$^{2}$. (One can notice that the conversion to $A/m^{2}$ can be easily achieved by multiplying $j_{\rm s}^{\rm eff}$ by the factor of $2e/\hbar$). We use the following definition of a spin resistance: $r_{\rm s}=\rho \SDL$.

\begin{figure}[hbt]
\includegraphics[width=12cm]{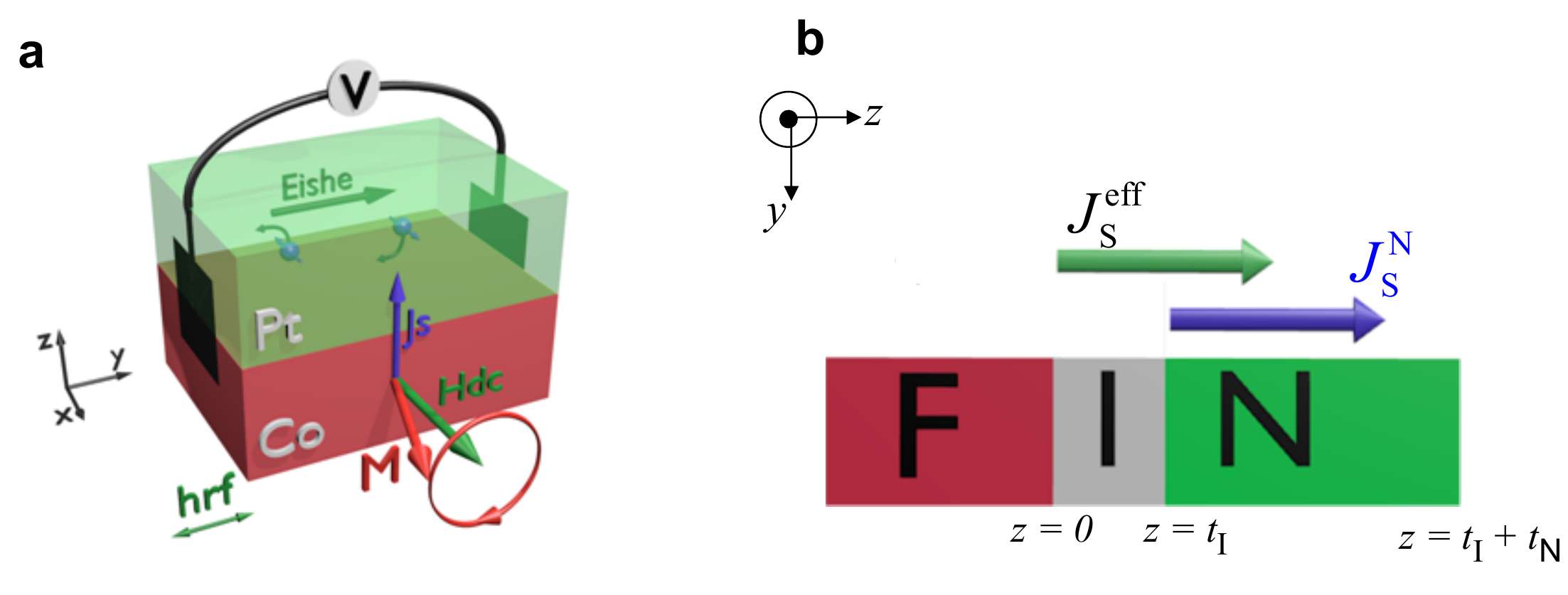}
\caption{Schematic representation of a bilayer Ferromagnetic (F)/Non-Magnetic (N) structure at resonance condition with spin current density losses taken into account at the interface (I) of equivalent thickness $t_{\rm I}$.}
\label{figS1}
\end{figure}

Our model considers an interface (I) in between the ferromagnetic (F) and the non-magnetic (N) layers, corresponding to a trilayer F$|$I$|$N system. In this approach the thickness of the interface is denoted as $t_{\rm I}$ and its spin resistance as $r_{\rm sI}$; the thickness of
Non-Magnetic material is denoted as $t_{\rm N}$ and its spin `bulk' resistance as $r_{\rm sN}^{\infty}$ (Fig. S1). The general solution of Eq. (1) in one dimension, along the $z$-axis in respect to presented coordinates, is:
\begin{equation}
\mu_{\rm sk}(z)=A_{\rm k}\exp(z/\SDL^{\rm k})+B_{\rm k}\exp(-z/\SDL^{\rm k}) \quad , \label{eq3}
\end{equation}
where each layer is indicated by the index $\rm k(=\rm I,N)$. The boundary conditions are the following ($\ ^{\prime}$ denotes the spacial derivative along $z$):
\begin{enumerate}
\item Chemical potential relation with the spin current: $\mu_{\rm sI}^{\prime}(z=0)=-ej_{\text{s0}}^{\text{eff}}\, r_{\rm sI}^{}/\SDL^{\rm I}$, spin current density is in A/m$^{2}$,
\item continuity of the electrochemical potential, $\mu_{\rm sI}(z=t_{\rm I})=\mu_{\rm sN}(z=t_{\rm I})$,
\item continuity of the spin current, $\mu_{\rm sI}^{\prime}(z=t_{\rm I})\SDL^{\rm I}/r_{\rm sI}^{}=\mu_{\rm sN}^{\prime}(z=t_{\rm I})\SDL^{\rm N}/r_{\rm sN}^{}$,
\item and finally the spin current vanishing at N/air interface (or N/susbtrate if the stacking order is reversed), $\mu_{\rm sN}^{\prime}(z=t_{\rm I}+t_{\rm N})=0$.
\end{enumerate}
\noindent The solutions can then be written as:

\begin{eqnarray}\label{eq4mu}
{{\mu }_{\text{sI}}}(z)&=&ej_{\text{s0}}^{\text{eff}}{{r}_{\text{sI}}}\frac{r_{\text{sN}}^{\infty }\cosh \left[ \frac{{{t}_{\text{I}}}-z}{\ell _{\text{sf}}^{\text{I}}} \right]\cosh \left[ \frac{{{t}_{\text{N}}}}{\ell _{\text{sf}}^{\text{N}}} \right]+{{r}_{\text{sI}}}\sinh \left[ \frac{{{t}_{\text{I}}}-z}{\ell _{\text{sf}}^{\text{I}}} \right]\sinh \left[ \frac{{{t}_{\text{N}}}}{\ell _{\text{sf}}^{\text{N}}} \right]}{r_{\text{sN}}^{\infty }\cosh \left[ \frac{{{t}_{\text{N}}}}{\ell _{\text{sf}}^{\text{N}}} \right]\sinh \left[ \frac{{{t}_{\text{I}}}}{\ell _{\text{sf}}^{\text{I}}} \right]+{{r}_{\text{sI}}}\cosh \left[ \frac{{{t}_{\text{I}}}}{\ell _{\text{sf}}^{\text{I}}} \right]\sinh \left[ \frac{{{t}_{\text{N}}}}{\ell _{\text{sf}}^{\text{N}}} \right]}\nonumber \\
{{\mu
}_{\text{sN}}}(z)&=&ej_{\text{s0}}^{\text{eff}}{{r}_{\text{sI}}}\frac{r_{\text{sN}}^{\infty
}\cosh \left[ \frac{{{t}_{\text{I}}}+{{t}_{\text{N}}}-z}{\ell
_{\text{sf}}^{\text{N}}} \right]}{r_{\text{sN}}^{\infty }\cosh
\left[ \frac{{{t}_{\text{N}}}}{\ell _{\text{sf}}^{\text{N}}}
\right]\sinh \left[ \frac{{{t}_{\text{I}}}}{\ell
_{\text{sf}}^{\text{I}}} \right]+{{r}_{\text{sI}}}\cosh \left[
\frac{{{t}_{\text{I}}}}{\ell _{\text{sf}}^{\text{I}}} \right]\sinh
\left[ \frac{{{t}_{\text{N}}}}{\ell _{\text{sf}}^{\text{N}}}
\right]}\quad ,
\end{eqnarray}

\noindent where $\mu_{\rm sI}(z)$ is valid when: $0\leq z\leq t_{\rm I}$; and $\mu_{\rm sN}(z)$ when: $t_{\rm I}\leq z\leq t_{\rm I}+t_{\rm N}$. Note that in similar way $\mu_{\rm sF}(z)$ can be obtained  for the profile inside F layer. However this calculation is not needed since we are already considering the backflow spin current density inside $j_{\text{s0}}^{\text{eff}}$ [Fig. \ref{figS1}(b)]. We can thus obtain the spin current profile along each layer:

\begin{eqnarray}\label{eq5Js(z)}
{{j}_{\text{sI}}}(z)&=&j_{\text{s0}}^{\text{eff}}\frac{r_{\text{sN}}^{\infty
}\sinh \left[ \frac{{{t}_{\text{I}}}-z}{\ell
_{\text{sf}}^{\text{I}}} \right]\cosh \left[
\frac{{{t}_{\text{N}}}}{\ell _{\text{sf}}^{\text{N}}}
\right]+{{r}_{\text{sI}}}\cosh \left[ \frac{{{t}_{\text{I}}}-z}{\ell
_{\text{sf}}^{\text{I}}} \right]\sinh \left[
\frac{{{t}_{\text{N}}}}{\ell _{\text{sf}}^{\text{N}}}
\right]}{r_{\text{sN}}^{\infty }\cosh \left[
\frac{{{t}_{\text{N}}}}{\ell _{\text{sf}}^{\text{N}}} \right]\sinh
\left[ \frac{{{t}_{\text{I}}}}{\ell _{\text{sf}}^{\text{I}}}
\right]+{{r}_{\text{sI}}}\cosh \left[ \frac{{{t}_{\text{I}}}}{\ell
_{\text{sf}}^{\text{I}}} \right]\sinh \left[
\frac{{{t}_{\text{N}}}}{\ell _{\text{sf}}^{\text{N}}} \right]}
\nonumber \\
{{j}_{\text{sN}}}(z)&=&j_{\text{s0}}^{\text{eff}}\frac{{{r}_{\text{sI}}}\sinh
\left[ \frac{{{t}_{\text{I}}}+{{t}_{\text{N}}}-z}{\ell
_{\text{sf}}^{\text{N}}} \right]}{r_{\text{sN}}^{\infty }\cosh
\left[ \frac{{{t}_{\text{N}}}}{\ell _{\text{sf}}^{\text{N}}}
\right]\sinh \left[ \frac{{{t}_{\text{I}}}}{\ell
_{\text{sf}}^{\text{I}}} \right]+{{r}_{\text{sI}}}\cosh \left[
\frac{{{t}_{\text{I}}}}{\ell _{\text{sf}}^{\text{I}}} \right]\sinh
\left[ \frac{{{t}_{\text{N}}}}{\ell _{\text{sf}}^{\text{N}}}
\right]}\quad .
\end{eqnarray}

\noindent The above equations confirm that $j_{\rm sI}(0)=j_{\text{s0}}^{\text{eff}}$, $j_{\rm sI}(t_{\rm I})=j_{\rm sN}(t_{\rm I})\equiv J_{\rm S}^{\rm N}$, and the ratio $R_{\rm SML}$between spin currents at each interface can be obtained:

\begin{equation}
{{R}_{\text{SML}}}\equiv \frac{J_{\text{S}}^{\text{N}}}{J_{\text{S}}^{\text{eff}}}\equiv \frac{{{j}_{\text{sN}}}({{t}_{\text{I}}})}{{{j}_{\text{sI}}}(0)}=\frac{{{r}_{\text{sI}}}}{{{r}_{\text{sI}}}\cosh \left[ \frac{{{t}_{\text{I}}}}{\ell _{\text{sf}}^{\text{I}}} \right]+r_{\text{sN}}^{\infty }\coth \left[ \frac{{{t}_{\text{N}}}}{\ell _{\text{sf}}^{\text{N}}} \right]\sinh \left[ \frac{{{t}_{\text{I}}}}{\ell _{\text{sf}}^{\text{I}}} \right]} \quad ,
\label{eq6ratioj2j1}
\end{equation}

\noindent which is the same expression given in the main text while taking into account the spin memory loss parameter $\delta=t_{\rm
I}/\SDL^{I}$.


\subsection*{Charge current: the correction factor due to the spin memory loss }

Now using the Eq. (\ref{eq5Js(z)}) we will show that the correction factor in the dc charge current due to the inverse spin Hall effect (ISHE) is equal to the ratio $R_{\rm SML}$ given in Eq. (\ref{eq6ratioj2j1}). In this approach we consider that the SHE is present only in the N layer and not at the interface layer. For that we reformulate the spin-to-charge conversion in the spin pumping-ISHE model\cite{SupMosendz2010c}:

\begin{equation}
\overrightarrow{j}_{\rm C}^{\rm ISHE}(z)=\theta_{\rm SHE}\, j_{\rm s}(z)\,[\overrightarrow{n}\times\overrightarrow{s}]\quad ,
\label{eq7ISHE1}
\end{equation}
\noindent where $\theta_{\rm SHE}$ is the spin Hall angle, $\overrightarrow{n}$ is the unit vector normal to the interface (along $z$ in used coordinates), and $\overrightarrow{s}$ is the spin polarization in $j_{\rm {s}}(z)$ which is parallel to the  magnetization at equilibrium in F layer (along $x$). The dc electric field created to compensate such induced charge current is then directed along $y$. Taking into account the shunting effect in the F layer, the total charge current writes: 
\begin{equation}
\int_{-t_{\rm F}}^{t_{\rm I}+t_{\rm N}}\left(j_{\rm C}^{\rm ISHE}(z)+\text{\ensuremath{\sigma}E}_{y}\right)\,dz=0\quad ,
\label{eq8jc:0}
\end{equation}

\noindent where $\sigma$ is the conductivity. While considering existence of the SHE only in the N layer the Eq. (\ref{eq8jc:0}) becomes:

\begin{equation}
\left(\sigma_{\rm F}t_{F}+\sigma_{\rm I}t_{i}+\sigma_{\rm N}t_{\rm N}\right)E_{y}=-\theta_{\rm {SHE}}^{\rm N}\int_{t_{\rm I}}^{t_{\rm I}+t_{\rm N}}j_{\rm sN}(z)dz\quad ,
\label{eq9Ey:js2}
\end{equation}

\noindent where $j_{\rm sN}(z)$ is expressed in A/m$^{2}$ (following units of $j_{\text{s0}}^{\text{eff}}$). Note that $j_{\rm sF}(z)$ is not
considered explicitly \cite{SupJiao2013} since the backflow spin current is already taken into account by the term $j_{\text{s0}}^{\text{eff}}$. Note also that $(\sigma_{\rm F}t_{\rm F}+\sigma_{\rm I}t_{\rm I}+\sigma_{\rm N}t_{\rm N})=R_{\rm sh}^{-1}$, is an inverse of a sheet resistance of the full stack multilayer. In practice one measures the dc voltage $V$ ($=E_{y}L$) along the length $L$ of the sample, but the physical parameter taken for the analysis is the charge current, $I_{\rm C}$. The charge current is determinate experimentally by the normalization: $I_{\rm C}=V/R$, with the resistance of the sample of width \emph{W} being: $R=R_{\rm sh}L/W$. On the other hand the charge current can also be expressed as: $I_{\rm C}=E_{y}L/R=E_{y}W/R_{\rm sh}$ When following Eq. (\ref{eq9Ey:js2}) this leads to:

\begin{equation}
I_{\rm C}=-\theta_{\text{SHE}}^{\rm N}\,W\int_{t_{\rm I}}^{t_{\rm I}+t_{\rm N}}j_{\rm sN}(z)dz\quad .
\label{eq10Ic:js2}
\end{equation}

Then using Eq. (5) one finds:
\begin{equation}
{{I}_{\text{C}}}=-W \theta
_{\text{SHE}}^{\text{N}}\ell
_{\text{sf}}^{\text{N}}\tanh \left[ \frac{{{t}_{\text{N}}}}{2\ell
_{\text{sf}}^{\text{N}}}\right] J_{\text{s0}}^{\text{eff}}\frac{{{r}_{\text{sI}}}}{{{r}_{\text{sI}}}\cosh \left[
\frac{{{t}_{\text{I}}}}{\ell _{\text{sf}}^{\text{I}}}
\right]+r_{\text{sN}}^{\infty }\coth \left[
\frac{{{t}_{\text{N}}}}{\ell _{\text{sf}}^{\text{N}}} \right]\sinh
\left[ \frac{{{t}_{\text{I}}}}{\ell _{\text{sf}}^{\text{I}}}
\right]} \quad .
\label{eq11Ic:jsSML}
\end{equation}

Here we observe the new factor in charge current, $R_{\rm SML} = \frac{j_{\rm sN}(t_{\rm I})}{j_{\rm sI}(0)}$, additionally to the usual $\SDL^{\rm N}\tanh\left[\frac{t_{\rm N}}{2\SDL^{\rm N}}\right]$ dependence used in most of the spin pumping ISHE studies. We point out that this new term is due to the spin memory loss at interface between F and N layers. Note that Eq. (\ref{eq11Ic:jsSML}) can be generalized for any multilayer structure with the SHE attributed to each one or only some of the layers. We point out that the sign `$-$' in Eq. (\ref{eq11Ic:jsSML}) means in our convention of Fig.~\ref{figS1}(a) negative voltage peak measured in $\|$F$|$M stacking order for a N material with positive $\theta_{SHE}$. This sign changes if we reverse the stacking order, the dc applied magnetic field or we turn the sample 180$^{\circ}$ around the $y$-axis. Indeed, as we have shown in Fig. 2(c), we observe negative voltage for our $\|$Co$|$Pt system and then it changes its sign when sample is turned 180$^{\circ}$.


\subsection{Effective spin mixing conductivity $\gupdw_{\rm eff}$ in F$|$I$|$N system }

According to the spin pumping theory, in the limit $\omega\ll1/\tau_{\rm sf}^{\rm N}$ and neglecting imaginary part of spin mixing conductivity, the effective spin mixing conductivity writes~\cite{SupTserkovnyak2005,SupAzevedo2011,SupNakayama2012,SupJiao2013}:
\begin{equation}
g_{\text{eff}}^{\uparrow \downarrow }=\frac{g_{{}}^{\uparrow \downarrow }}{1+\tilde{g}_{{}}^{\uparrow \downarrow }\beta }\quad .
\label{geff1}
\end{equation}

\noindent Here $\tilde{g}^{\uparrow \downarrow }$ satisfy $\frac{2{{e}^{2}}}{h}{{\tilde{g}}^{\uparrow \downarrow }}r_{\text{sN}}^{\infty }=\frac{1}{2}\sqrt{\frac{3}{\varepsilon }}$ and $\varepsilon=\tau_{\rm el}/\tau_{\rm sf}^{\rm N}$ is the ratio of the spin-conserved to spin-flip relaxation times. In pure bilayer with transparent interfaces the $\beta$ back flow factor writes $\beta=(2e^2/h)r_{\rm sN}^\infty \coth(t_{\rm N}/\ell_{\rm sf}^{\rm N})$. Now we can calculate the back flow $\beta$ factor in our F/I/N system according to $\beta=\frac{2e}{h}\frac{\mu_{\rm
sI}(0)}{j_{\text{s0}}^{\text{eff}}}$ \cite{SupTserkovnyak2005,SupHarii2012,SupJiao2013,SupBoone2013}. Then replacing $\mu_{\rm
sI}(0)$ by using Eq. (\ref{eq4mu}), $\beta$ becomes:
\begin{equation}\label{eq12Beta}
\beta
=\frac{2{{e}^{2}}}{h}{{r}_{\text{sI}}}\frac{r_{\text{sN}}^{\infty
}\cosh \left[ \frac{{{t}_{\text{I}}}}{\ell _{\text{sf}}^{\text{I}}}
\right]\cosh \left[ \frac{{{t}_{\text{N}}}}{\ell
_{\text{sf}}^{\text{N}}} \right]+{{r}_{\text{sI}}}\sinh \left[
\frac{{{t}_{\text{I}}}}{\ell _{\text{sf}}^{\text{I}}} \right]\sinh
\left[ \frac{{{t}_{\text{N}}}}{\ell _{\text{sf}}^{\text{N}}}
\right]}{r_{\text{sN}}^{\infty }\cosh \left[
\frac{{{t}_{\text{N}}}}{\ell _{\text{sf}}^{\text{N}}} \right]\sinh
\left[ \frac{{{t}_{\text{I}}}}{\ell _{\text{sf}}^{\text{I}}}
\right]+{{r}_{\text{sI}}}\cosh \left[ \frac{{{t}_{\text{I}}}}{\ell
_{\text{sf}}^{\text{I}}} \right]\sinh \left[
\frac{{{t}_{\text{N}}}}{\ell _{\text{sf}}^{\text{N}}} \right]}\quad ,
\end{equation}

\noindent which is equivalent to the $\beta$  factor shows in ref.
\cite{SupHarii2012} and \cite{SupBoone2013}. Using Eq. (\ref{eq12Beta}) and after simple mathematical manipulations:
\begin{equation}
g_{\text{eff}}^{\uparrow \downarrow }={{g}^{\uparrow \downarrow }}\frac{{{r}_{\text{sI}}}\cosh \left[ \frac{{{t}_{\text{I}}}}{\ell _{\text{sf}}^{\text{I}}} \right]+r_{\text{sN}}^{\infty }\coth \left[ \frac{{{t}_{\text{N}}}}{\ell _{\text{sf}}^{\text{N}}} \right]\sinh \left[ \frac{{{t}_{\text{I}}}}{\ell _{\text{sf}}^{\text{I}}} \right]}{{{r}_{\text{sI}}}\left( 1+\frac{1}{2}\sqrt{\frac{3}{\varepsilon }}\coth \left[ \frac{{{t}_{\text{N}}}}{\ell _{\text{sf}}^{\text{N}}} \right] \right)\cosh \left[ \frac{{{t}_{\text{I}}}}{\ell _{\text{sf}}^{\text{I}}} \right]+\left( r_{\text{sN}}^{\infty }\coth \left[ \frac{{{t}_{\text{N}}}}{\ell _{\text{sf}}^{\text{N}}} \right]+\frac{1}{2}\frac{{{r}_{\text{sI}}}^{2}}{r_{\text{sN}}^{\infty }}\sqrt{\frac{3}{\varepsilon }} \right)\sinh \left[ \frac{{{t}_{\text{I}}}}{\ell _{\text{sf}}^{\text{I}}} \right]}\quad .
 \label{eq14JsefJs0b}
\end{equation}

\begin{figure}[hbt]
\includegraphics[width=10cm]{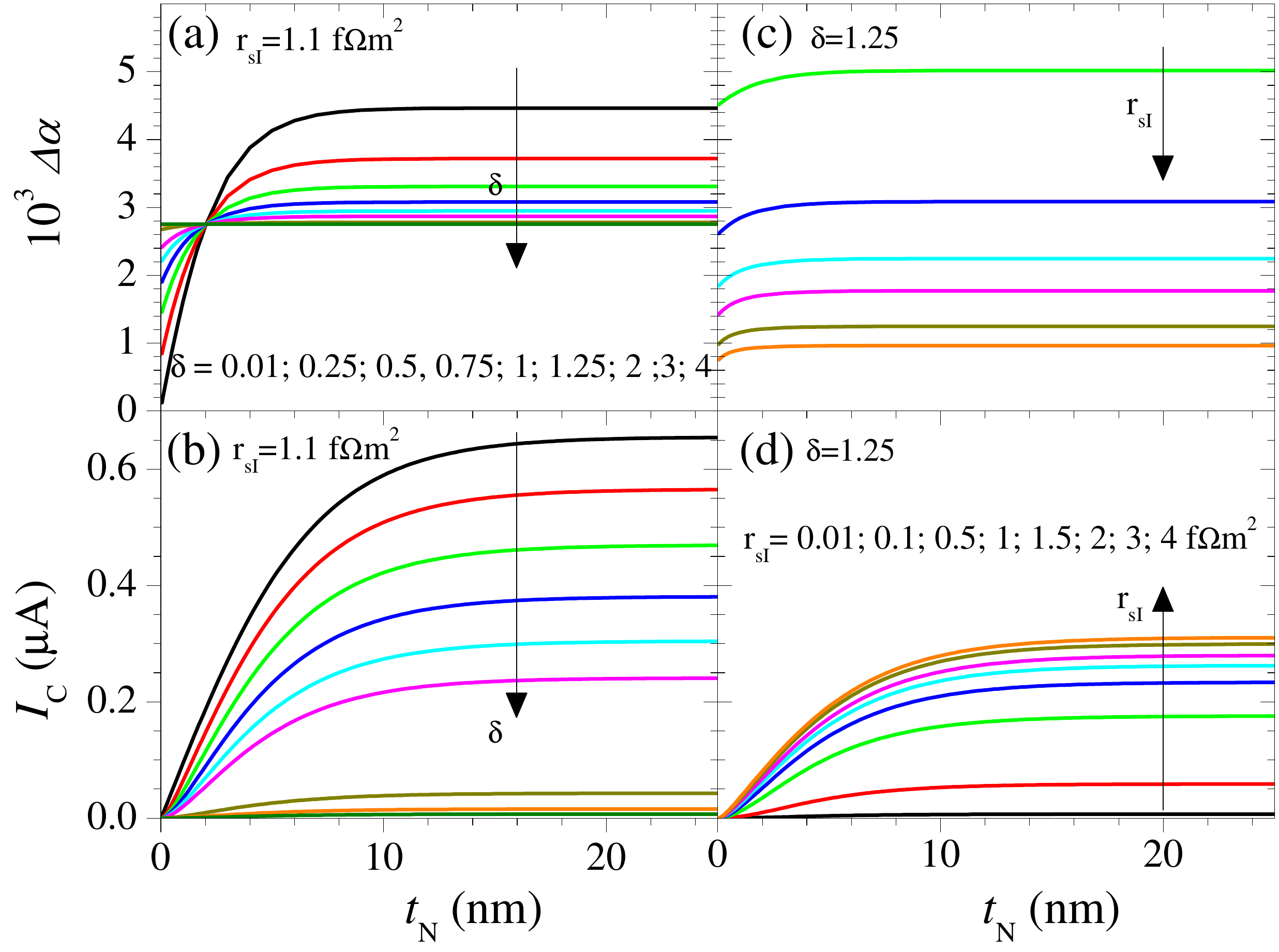} 
\caption{Simulated curves of $t_{\rm N}$ dependence of: (a,c) damping constant enhancement according to Eq. (\ref{eq15Dalfa}), and (b,d) charge current following Eq. (\ref{eq11Ic:jsSML}). Most of the parameters used correspond to our $\|$Co(15)$\vert$Cu(5)$\vert$Pt(t) system. For (a,c): $M_{\rm eff}=1330$~emu/cm$^3$, $t_{\rm F}=15$~nm, $\varepsilon=0.1$, and $\SDL^{\rm N}=3.4$~nm. For (b,d): along with the same $\SDL^{\rm N}$, $J_{\rm S}^{\rm eff}=9.5$~MA/m$^2$ $\SHA^{\rm N}=0.051$, and $W=0.4$~mm. The
black arrows indicate the increasing sense of $\delta$ (a,c) or $r_{\rm sI}$ (b,d) for the values show in each panel. The curves for $\delta=0.01,0.1$ in (c) are out of scale.}
\label{figS2}
\end{figure}

\subsection*{Determination of enhancement damping constant}
We rewrite the enhancement of the damping constant neglecting the imaginary part of spin mixing conductivity, \cite{SupTserkovnyak2003,SupBoone2013}:
\begin{eqnarray}
  \Delta \alpha &=&\frac{g{{\mu }_{\text{B}}}}{4\pi {{M}_{\text{eff}}}{{t}_{\text{F}}}}g_{\text{eff}}^{\uparrow \downarrow }\nonumber\\
  &=&\frac{g{{\mu }_{\text{B}}}}{4\pi {{M}_{\text{eff}}}{{t}_{\text{F}}}}{{g}^{\uparrow \downarrow }}\frac{{{r}_{\text{sI}}}\cosh \left[ \frac{{{t}_{\text{I}}}}{\ell _{\text{sf}}^{\text{I}}} \right]+r_{\text{sN}}^{\infty }\coth \left[ \frac{{{t}_{\text{N}}}}{\ell _{\text{sf}}^{\text{N}}} \right]\sinh \left[ \frac{{{t}_{\text{I}}}}{\ell _{\text{sf}}^{\text{I}}} \right]}{{{r}_{\text{sI}}}\left( 1+\frac{1}{2}\sqrt{\frac{3}{\varepsilon }}\coth \left[ \frac{{{t}_{\text{N}}}}{\ell _{\text{sf}}^{\text{N}}} \right] \right)\cosh \left[ \frac{{{t}_{\text{I}}}}{\ell _{\text{sf}}^{\text{I}}} \right]+\left( r_{\text{sN}}^{\infty }\coth \left[ \frac{{{t}_{\text{N}}}}{\ell _{\text{sf}}^{\text{N}}} \right]+\frac{1}{2}\frac{{{r}_{\text{sI}}}^{2}}{r_{\text{sN}}^{\infty }}\sqrt{\frac{3}{\varepsilon }} \right)\sinh \left[ \frac{{{t}_{\text{I}}}}{\ell _{\text{sf}}^{\text{I}}} \right]}\quad ,
\label{eq15Dalfa}
\end{eqnarray}
where $g$ is the Land$\acute{e}$ g-factor (of the F layer), $M_{\text{eff}}$ is the effective magnetic saturation of F, and $\mu_{B}$ is the Bohr magnetron. In the case of Pt, $\varepsilon=0.1$ ~\cite{SupTserkovnyak2005,SupAzevedo2011,SupNakayama2012,SupJiao2013}

We see now from Eqs. (\ref{eq11Ic:jsSML}), (\ref{eq14JsefJs0b}) and (\ref{eq15Dalfa}) that the enhancements of the damping constant and the charge current have a different length scale dependence on the thickness of the N material due to the spin memory losses at the interface. The $t_{\rm N}$ dependence of damping enhancement $\Delta \alpha$ with different parameters $\delta$ is shown in Fig. \ref{figS2}(a), and $r_{\rm SI}$ in Fig. \ref{figS2}(c). The same dependence on $I_{\rm C}$ [Fig. \ref{figS2}(b,d)] clearly shows the $\SDL^{\rm N}$ lengthscale; and such lengthscale does not change for any set of $\delta$ or $r_{\rm sI}$ parameters. In Fig. \ref{figS2}(a) all
the curves are intersected at $t_{\rm N}=\SDL^{\rm N}/2$ and for smaller thickness the curve change strongly only when $\delta\lsim 0.25$. In Fig. \ref{figS2}(b) we can observe that the saturation level of charge current is quickly reduced with enhancement of $\delta$. As consequence, large $\delta$ parameter will quickly increase the damping constant with no charge current production [Fig. \ref{figS2}(a,b)]. It happens the opposite tendency with the $r_{\rm sI}$ parameter: very small values increase the damping constant without charge current production [Fig. \ref{figS2}(c,d)]. The saturation level of charge current does not change significantly for $r_{\rm sI}\gsim 3$.

\subsection{Derivation of the spin resistance and spin memory loss parameters in the multi-layer case.}
In Co$|$Pt systems, the complete analysis of the profile of the spin-current pumped from Co and dissipated both at the interface including spin memory loss (SML) and in the `bulk' Pt heavy metal requires a three-layer treatment. In a diffusive approach, the transport of the longitudinal component of the spin current is parametrized by the spin-resistance $r_{sF}$ of the ferromagnet, the spin-resistance of the thin interface layer $r_{sI}=r_b/\delta$ and the one, $r_{sN}$, of the heavy metal (Pt). The interface extends on a scale of a few units of atomic planes (fraction of nanometer) corresponds to local magnetic fluctuations and disorder responsible for partial spin depolarization and spin-current discontinuities. It is generally characterized by its characteristic resistance $r_b$ and the SML parameter $\delta=t_I/\ell_{sf}^I$ which can be viewed as the ratio between the effective interface thickness ($t_I$) and the corresponding interfacial spin diffusion length ($\ell_{sf}^I$). Hereafter, we note $P_\infty$ the bulk spin asymmetry coefficient of the ferromagnet, however not relevant in the mechanism of spin-pumping and related spin-current diffusion and disregard the interfacial spin-asymmetry coefficient $\gamma$.

Using the transfer matrix method \cite{SupJaffres2010} adapted to the longitudinal spin-current propagation in magnetic multilayers within a diffusive approach, one can calculate the current spin-polarization ($\mathcal{P}$) at each side of the Co$|$I$|$Pt interface at the respective Co side ($\mathcal{P}^{in}_{(F)}$) and Pt ($\mathcal{P}^{out}_{(N)}$) sides according to :

\begin{eqnarray}
\mathcal{P}^{in}_{(F)}=\frac{P_\infty [1+\frac{r_{sN}\delta}{r_b}~\sinh(\delta)]}{[1+\frac{r_{sN}}{r_{sF}}]\cosh(\delta)+\frac{r_b}{r_{sF}}\frac{\sinh(\delta)}{\delta}+\frac{r_{sN} \delta}{r_b}~\sinh(\delta)}\\
\mathcal{P}^{out}_{(N)}=\frac{P_\infty}{[1+\frac{r_{sN}}{r_{sF}}]~\cosh(\delta)+\frac{r_b}{r_{sF}}\frac{\sinh(\delta)}{\delta}+\frac{r_{sN} \delta}{r_b} \sinh(\delta)}
\end{eqnarray}
showing up the spin-current discontinuity (or spin-memory loss) with a probability of spin-conserving of the order of $\exp(-\delta)$ between (F) and (N). The two terms in the denominator, $\frac{r_{sN}}{r_{sF}}~\cosh(\delta)$ and $\frac{r_b}{r_{sF}}~\frac{\sinh(\delta)}{\delta}$, describe the impedance mismatch issue at the `left side' of the interface impeding an efficient injection of a spin-polarized current from Co into a non-magnetic highly resistive bulk material (first term) or into a highly resistive interface (second term). On the other hand, the last term $\frac{r_{sN}~\delta}{r_b}~\sinh(\delta)$ describes the impedance mismatch issue at the `right side' of the interface impeding an efficient injection if the interfacial spin-resistance is too small and then responsible of supplementary spin-flip processes by spin-backflow processes from the `right side'.

The ratio betweeen \textit{in} and \textit{out} spin-current can be more simply expressed as:
\begin{eqnarray}
\zeta=\frac{\mathcal{P}^{out}_{(N)}}{\mathcal{P}^{in}_{(F)}}=\frac{1}{\cosh(\delta)+\frac{r_{sN}}{r_I}\sinh(\delta)} \quad ,
\end{eqnarray}
which solely depends, by renormalization, on the interface property on the `right', and not of the spin-resistance of the ferromagnetic injector (Co). $\zeta$ quantifies the ratio between the rate of spin-flips in the interface region ($I$) itself scaling like $\delta/r_b$ to the rate of spin-flips inside the spin-sink material ($N$) scaling like $1/r_{sN}$. This ratio is the same as the one calculated previously in Eq.~\ref{eq6ratioj2j1}. Note that, in the limit of large $\delta$, the latter expression transforms into $\zeta\simeq \frac{1}{1+\frac{\delta r_{sN}}{r_b}}~\exp(-\delta)$. Thus, apart from the expected exponential decrease for a short SDL within the thin interfacial region, the spin-polarized current penetrating $N$ strongly depends on $r_{sN}\delta /r_b=r_{sN}/r_{sI}$ from the argument of impedance mismatch at the right hand side of the interface. The ensemble of arguments developed here to find the dominant spin-flip contribution between interface region and outward material $N$ can now be applied to treat simply the case of two (or several) SML interfaces placed in series as discussed now.

In the case of Co$|$Cu$|$Pt trilayers involving spin-memory loss (SML) at both Co$|$Cu and Cu$|$Pt interface, a five-layer model is the more generally needed for the calculation of the spin-polarized current profile throughout the structure. However, in absence of any spin-flips in the Cu spacer (because of its long spin diffusion length compared to its thickness), a three-layer treatment becomes possible if the two consecutive SML interfaces are treated like a single effective SML one. The following calculations generalizes these idea by considering the spin-current injected at the level of the two consecutive SML interfaces neglecting the Cu spacer. One then notes $r_{b,i}$, $\delta_i$ and $r_{sI},i=r_{b,i}/\delta_i$, the interface resistance, spin-memory loss (SML) parameter and effective spin-resistance of the respective Co$|$Cu $(i=1)$ and Cu$|$Pt ($i=2$) interfaces. Following the previous arguments and in the limit of a large SML within the second interface ($\delta_2>1$), one can calculate respectively the spin-current $\mathcal{P}_{\rm Cu}$ injected in the Cu spacer (between the two SML interfaces) and the one penetrating the Pt sink $\mathcal{P}_{N}$ according to:

\begin{eqnarray}
\mathcal{P}_{\rm Cu}\approx\frac{\mathcal{P}_{in}}{\cosh(\delta_1)+\frac{r_{sI,2}}{r_{sI,1}}\sinh(\delta_1)}\\
\mathcal{P}_{N}\approx\frac{\mathcal{P}_{\rm Cu}}{\cosh(\delta_2)+\frac{r_{sN}}{r_{sI,2}}\sinh(\delta_2)}
\end{eqnarray}

Describing the two SML interfaces in series by a single effective one characterized by $r_{b}^{\rm eff}$, $\delta^{\rm eff}$ and $r_{sI}^{\rm eff}=r_{b}^{\rm eff}/\delta^{\rm eff}$) with
\begin{eqnarray}
\mathcal{P}_{N}\approx \frac{\mathcal{P}_{in}}{\cosh(\delta^{\rm eff})+\frac{r_{sN}}{r_{sI}^{\rm eff}}\sinh(\delta^{\rm eff})}
\end{eqnarray}
leads to the determination of $\delta^{\rm eff}$ and $r_{sI}^{\rm eff}=r_{b}^{\rm eff}/\delta^{\rm eff}$ by matching the two solutions according to:
\begin{eqnarray}
\nonumber \cosh(\delta^{\rm eff})=\cosh(\delta_2)~ \left[ \cosh(\delta_1)+\frac{r_{sI,2}}{r_{sI,1}}\sinh(\delta_1) \right ]\\
\frac{1}{r_{sI}^{\rm eff}}=\frac{\sinh(\delta_2)}{\sinh(\delta^{\rm eff})}~ \left[ \frac{1}{r_{sI,2}}\cosh(\delta_1)+\frac{1}{r_{sI,1}}\sinh(\delta_1) \right]
\label{EqEffectiveDelta}
\end{eqnarray}
The table \ref{tabledeltaRi} displays the literature and calculated values for the SML parameters. We used the table parameters which gave us the SHA values of $5.7\pm0.3$\% and $5.6\pm0.1$\% for the trilayer and the bilayer respectively. Using a combined fitting procedure, one finds $\SHA^{\rm Pt}=5.6\pm 0.1$\% (statistical error only).

\begin{table}[htb]
\caption{\label{tabledeltaRi} Values of $\delta$, $r_b=AR^\star$ for F$|$N interfaces ($r_b=2AR$ for N$_1|$N$_2$ interfaces) and $r_{sI}$ as reported in the literature (at 4.2\,K), or calculated with Eq. \ref{EqEffectiveDelta}}
\begin{tabular}{l c c c c c}
System	&		$\quad \delta\quad $	&	$\quad 2AR^\star$\,(f$\Omega$m$^2$)\quad 	&	$\quad 2AR$\,(f$\Omega$m$^2$)\quad 	&	$\quad r_{sI}$\,(f$\Omega$m$^2$)\quad &	Ref.	\\
\hline
\hline
Co$|$Cu	&	0.25	&	1.0		& --	&	2.0		&	\cite{SupEid2002} 			\\
Cu$|$Pt	&	0.9		&	--		&	1.5	&	1.7		&	\cite{SupKurt2002}			\\
Co$|$Pt	&	0.9		&	1.5		&	--	&	0.83		&	\cite{SupNguyen2013}		\\
Co$|$Cu$|$Pt\quad	&	1.2	&	2.0	&	--	&	0.85	&	calculation 
\end{tabular}
\end{table}

\subsection*{Frequency dependence}
\begin{figure}[hbt]
\includegraphics[width=12cm]{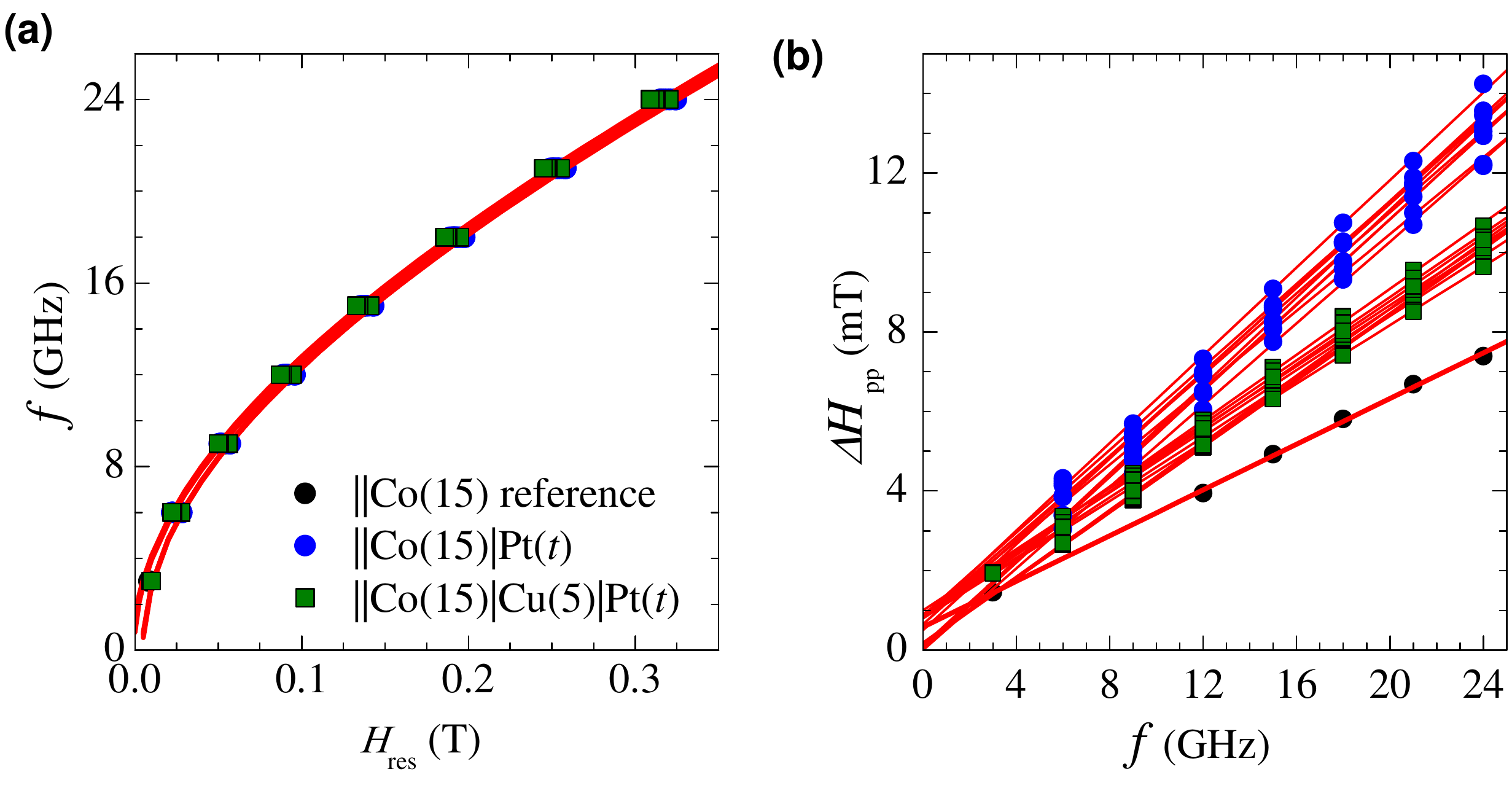} \caption{(a) Frequency \emph{vs}. resonance field and
its fitting (red line) according to Eq. (\ref{eqKittel}). (b)
Peak-to-peak linewidth \emph{vs}. frequency and its linear fit (red
line) following Eq. (\ref{eqDHvsf}) for the $\|$Co(15)$\vert$Al($7$)
reference sample, the $\|$Co(15)$\vert$Pt($t$) series and the
$\|$Co(15)$\vert$Cu$(5)\vert$Pt($t$) series.} \label{figS3}
\end{figure}

We have measured the FMR spectrum at different frequencies in order to determine the effective saturation magnetization as well as the damping constant $\alpha$. This experiment was performed in a broadband ($3-24$~GHz) using a strip-line antenna and a vector network analyzer (VNA). The frequency $f$ {\it vs.} the magnetic resonance field and the linewidth ($\Delta H_{pp}$) {\it vs.} $f$ for the in-plane configuration are displayed in Figure  \ref{figS3}. By this method, we have also evaluated the in-plane anisotropies and the inhomogeneous contributions to the FMR linewidth ($\Delta H_0$) according to the following relationships:
\begin{equation}
\label{eqKittel}
{{\left( \frac{\omega }{\gamma }\right)}^{2}}=(H+{{H}_{K}})(H+{{H}_{K}}+4\pi {{M}_{\text{eff}}})
\end{equation}

\begin{equation}
\label{eqDHvsf}
\Delta {{H}_{\text{pp}}}=\Delta{{H}_{\text{0}}}+\frac{2}{\sqrt{3}}\left( \frac{\omega }{\gamma }\right)\alpha
\end{equation}

\subsection*{Experimental determination of Co resistivity: Co thickness dependence of sheet resistance}
\begin{figure}[hbt]
\includegraphics[]{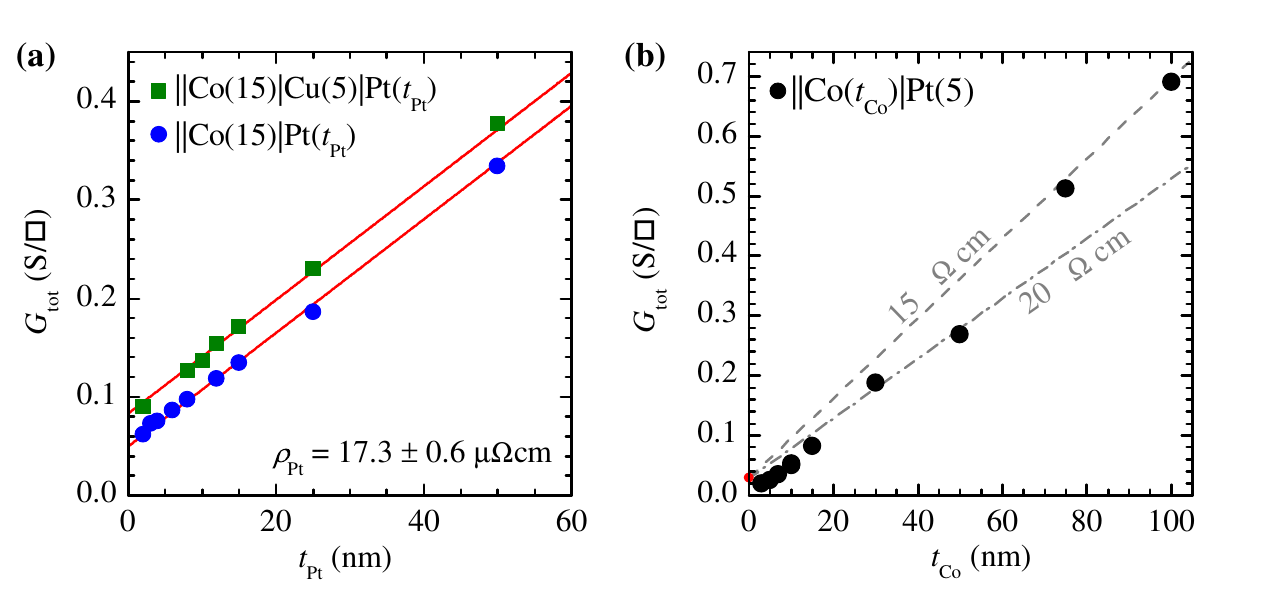} 
\caption{ (a) Pt thickness dependence of the total sheet conductance in $\|$Co(15)$|$Pt(t) and $\|$Co(15)$|$Cu(5)|Pt(t). Red lines represent the best combined linear fits from which Pt resistivity is evaluated. (b) Co thickness dependence of the total sheet conductance in $\|$Co(t)$|$Pt(5) films. Two slopes corresponding to two conductivities are indicated. The red dot is the expected 5\,nm-thick Pt sheet conductance.}
\label{figS4}
\end{figure}

As shown in the main text for the Pt resistivity, we have also measured the sheet resistance by 4 probes methods in $\|$Co(t)$\vert$Pt(15) samples. We show both, Pt and Co thickness dependence in Fig. \ref{figS4}. The linear fits are made according to: (i) $G_{{\rm Co}\vert{\rm Pt}}=G_{\rm 01}+\sigma_{\rm Pt}\cdot t_{Pt}$ for Pt thickness dependence where $\sigma_{\rm Pt}$ is the Pt conductivity and $G_{\rm 01}$ would be ideally the Co sheet conductance contribution. (ii) Similarly $G_{{\rm Co}\vert{\rm Pt}}=G_{\rm 02}+\sigma_{\rm Co}\cdot t_{Co}$ for Co thickness dependence. If one uses the $G_{01}$ value it gives an apparently Co resistivity of 29.1 $\mu\Omega$cm. However the Co thickness dependence displayed in Fig. \ref{figS4}(b) shows a more complex behavior. Details of this dependence are irrelevant for our study because the Co layer is deposited first and its thickness is kept fixed at 15\,nm. Nevertheless, one can note that the non-linearity of $G(t_{\rm Co})$ reveal a Co resistivity which is higher at low thickness, probably due to diffusion on the SiO$_2$ substrate surface. This phenomenon is not observed for the Pt thickness variation, in all likelihood because of the metallic Co `buffer'.

\end{document}